\documentclass[sort&compress,review]{elsarticle}
\usepackage{epsfig,amsmath,amssymb,euscript}
\usepackage{wasysym}
\usepackage{color}
\usepackage{accents}
\usepackage{hyperref}

\begin{document}

\begin{frontmatter}

\title{Electroweak effects in the extraction of the CKM angle~$\gamma$ from $B\to D \pi$ decays} 

\author{Joachim Brod} 
\address{PRISMA Cluster of Excellence and Mainz Institute for
  Theoretical Physics, Johannes-Gutenberg-Universit\"at, 55099
  Mainz, Germany} 
\ead{joachim.brod@uni-mainz.de} 

\begin{abstract}
The angle $\gamma$ of the standard CKM unitarity triangle can be
determined from tree-level $B$-meson decays essentially without
hadronic uncertainties. We calculate the second-order electroweak
corrections for the $B \to D \pi$ modes and show that their impact on
the determination of $\gamma$ could be enhanced by an accidental
cancellation of poorly known hadronic matrix elements. However, we do
not expect the resulting shift in $\gamma$ to exceed $\big|\delta
\gamma^{D\pi} /\gamma \big| \lesssim {\mathcal O}\big(10^{-4}\big)$.
\end{abstract}

\end{frontmatter}

\section{Introduction}

The Cabibbo-Kobayashi-Maskawa (CKM) angle $\gamma \equiv
\arg(-V_{ud}^{\phantom{*}} V_{ub}^*/V_{cd}^{\phantom{*}} V_{cb}^*)$
can be extracted from $B \to DK$ and $B \to D\pi$ decays that receive
contributions only from tree operators~\cite{Bigi:1981qs}. The absence
of penguin contributions and the fact that all relevant hadronic
matrix elements can be obtained from data makes this determination
theoretically extremely clean, thus providing a standard candle for
the search for physics beyond the standard model (SM).

\begin{figure}
  \centering
  \includegraphics[width=0.3\columnwidth]{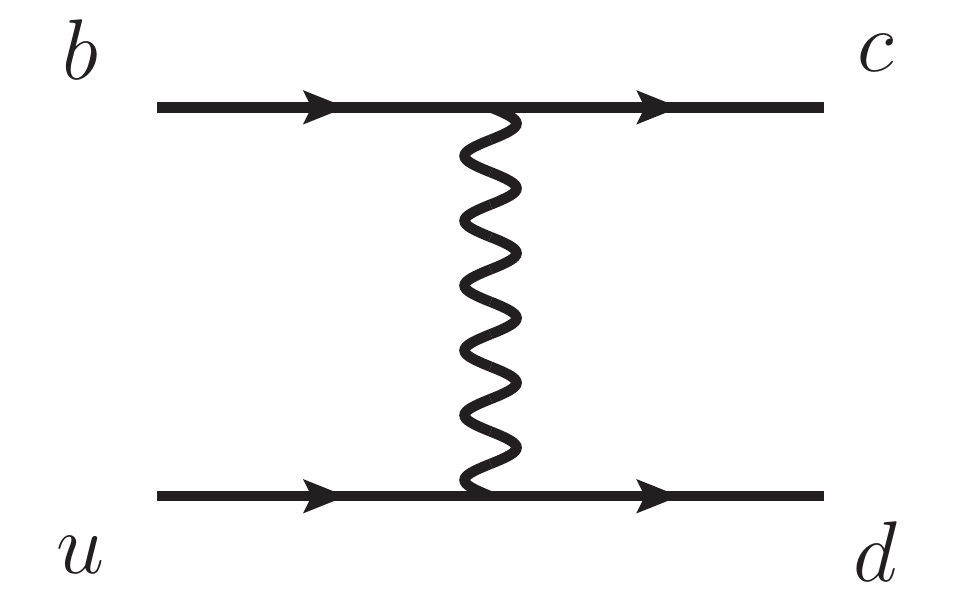}~~~~~~~~~~
  \includegraphics[width=0.3\columnwidth]{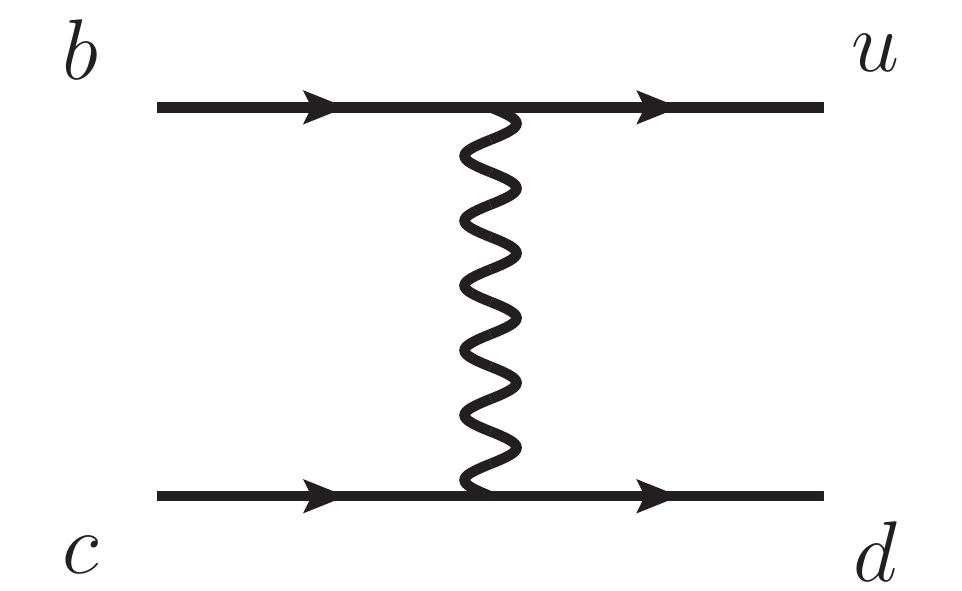}
  \caption{Tree contributions (with single $W$ exchange) that mediate
    $b\to c \bar u d$ (left) and $b\to u\bar c d$ (right) quark-level
    processes, which lead to $B^-\to D^0 \pi^-$ and $B^-\to \bar D^0
    \pi^-$ decays, respectively. }
\label{fig:tree}
\end{figure}

The sensitivity to $\gamma$ arises from the interference of $b\to
c\bar u q$ and $b\to u\bar c q$ decay amplitudes (see
Fig.~\ref{fig:tree}), which have a relative weak phase $\gamma$. Here,
$q$ denotes either a strange or a down quark. The quark-level
transitions with $q=s$ mediate the $B^- \to D^0 K^-$ and $B^- \to \bar
D^0 K^-$ decays, whereas the transitions with $q = d$ induce the $B^-
\to D^0 \pi^-$ and $B^- \to \bar D^0 \pi^-$ decays. In both cases the
$D^0$ and $\bar D^0$ mesons decay into a common final state $f$,
leading to the interference of the two decay channels. Several
variants of this method have been formulated, distinguished by the
final state $f$~\cite{Gronau:1990ra, Gronau:1991dp, Atwood:1996ci,
  Giri:2003ty, Grossman:2002aq, Bondar:2005ki}. Alternatively, one can
also use decays of neutral $B^0$ or $B_s^0$
mesons~\cite{Kayser:1999bu, Gronau:2004gt}, multibody $B$
decays~\cite{Aleksan:2002mh, Gershon:2009qr, Gershon:2008pe,
  Gershon:2009qc}, and $D^*$ or $D^{**}$ decays~\cite{Bondar:2004bi,
  Sinha:2004ct} (see also the reviews in~\cite{Zupan:2007zz}).

Whereas in most analyses $\gamma$ has been extracted only from the $B
\to DK$ modes, the LHCb collaboration recently included also the $B
\to D \pi$ modes in their full combination~\cite{Aaij:2013zfa,
  LHCbckm2014}. The sensitivity to $\gamma$ of these modes is smaller
than that of the $B \to D K$ modes, due to a smaller interference
term; this effect is, however, partially compensated by the larger $B
\to D \pi$ branching ratio.

The extraction of $\gamma$ from tree-level decays suffers from various
uncertainties. Some of them can be reduced once more statistics
becomes available, for instance, those related to a Dalitz-plot
analysis~\cite{Libby:2010nu, Grossman:2002aq, Aihara:2012aw,
  Aaij:2014uva}. Other sources of reducible uncertainties are $D-\bar
D$ mixing and, for final states with a $K_S$, also $K-\bar K$
mixing. Both of these effects can be taken into account by measuring
the mixing parameters and appropriately modifying the expressions for
the decay amplitudes~\cite{Silva:1999bd, Bondar:2010qs,
  Grossman:2005rp, Rama:2013voa}. In a similar manner, the effects of
nonzero $\Delta \Gamma_s$ can be included into the $\gamma$ extraction
from untagged $B_s\to D \phi$ decays~\cite{Gronau:2007bh}. It is also
possible to allow for CP violation in the $D$-meson
decays~\cite{Aaij:2013zfa, Wang:2012ie, Martone:2012nj,
  Bhattacharya:2013vc, Bondar:2013jxa}. The effects of CP violation in
kaon mixing have recently been discussed
in~\cite{Grossman:2013woa}. Finally, the impact on $\gamma$ of
new-physics contributions to tree-level Wilson coefficients has been
estimated in~\cite{Brod:2014bfa}.

As shown in~\cite{Brod:2013sga}, the first irreducible theory error on
the determination of $\gamma$ arises from higher-order electroweak
corrections. It has been calculated for the $B \to D K$ modes,
resulting in an upper bound on the shift in $\gamma$ of $\delta
\gamma^{DK}/\gamma\lesssim {\mathcal
  O}(10^{-7})$~\cite{Brod:2013sga}. The shift due to electroweak
corrections for the extraction of $\gamma$ from the $B \to D \pi$
modes has not yet been computed; we close the gap in this letter.

The main difference between the $B \to D \pi$ and the $B \to D K$
modes lies in their CKM structure. Consequently, as we will see later,
the effect of the electroweak corrections for the $B \to D \pi$ modes
could potentially be much larger than for the $B \to D K$ modes, due
to an approximate cancellation of hadronic matrix elements. However,
we do not expect the final shift in $\gamma$ to exceed $\big|\delta
\gamma^{D\pi}/\gamma \big| \lesssim {\mathcal O}\big(10^{-4}\big)$
without some accidental fine tuning -- well below the precision of any
current or future measurement.

This letter is organized as follows. In Sec.~\ref{sec:2} we calculate
the electroweak corrections to the relevant Wilson coefficients and
estimate the resulting shift in $\gamma$ in Sec.~\ref{sec:shift}. We
conclude in Sec.~\ref{Conclusions}.

\section{Calculation of the electroweak corrections}
\label{sec:2}

We calculate the shift in $\gamma$ due to electroweak corrections in
close analogy to the procedure in Ref.~\cite{Brod:2013sga}. The
sensitivity of the $B\to D\pi$ modes to $\gamma$ enters through the
amplitude ratio
\begin{eqnarray}\label{rB}
r_B^{D\pi} e^{i(\delta_B^{D\pi} - \gamma)} \equiv \frac{A(B^- \to \bar D^0
  \pi^-)}{A(B^- \to D^0 \pi^-)}, 
\end{eqnarray}
where $r_B^{D\pi} \in [0.001, 0.040]$ at 95\% CL~\cite{LHCbckm2014}
reflects the CKM and color suppression of the amplitude $A(B^-\to \bar
D^0 \pi^-)$ relative to the amplitude $A(B^- \to D^0 \pi^-)$ (there is
currently no constraint on the strong phase $\delta_B^{D\pi}$ at 95\%
CL). Note that the corresponding ratio $r_B^{DK}$ for the $B \to D K$
modes is known more precisely, $r_B^{DK} \in [0.0732, 0.1085]$ at 95\%
CL~\cite{LHCbckm2014}. Naive scaling by CKM factors leads to the
generic expectation $r_B^{D\pi} \approx 5 \times 10^{-3}$.

The equality~\eqref{rB} is valid only at leading order in the weak
interactions, ${\mathcal O}(G_F)$, where both the $b \to c \bar u d$
and $b\to u \bar c d$ transitions are mediated by a tree-level $W$
exchange\footnote{In Ref.~\cite{Grossman:2013woa} it has been pointed
  out that the weak phase entering the $B \to D K$ modes differs from
  $\gamma$ by subleading corrections of order $\lambda^4 \approx 2.6
  \times 10^{-3}$, where $\lambda \equiv |V_{us}|$ is the Wolfenstein
  parameter (note that in~\cite{Grossman:2013woa} $\lambda$
  erroneously appears raised to the power of 5). A similar observation
  applies for the $B \to D \pi$ modes. The relation of the phase of
  $r_B^{D\pi}$ to $\gamma$ involves the ratio $V_{cd}^2/V_{ud}^2 =
  \lambda^2 [1 - \lambda^4 A^2 (1 - 2(\rho + i \eta)) + {\mathcal
      O}(\lambda^6)]$. This introduces another small ${\mathcal
    O}(\lambda^4)$ uncertainty into the extraction of $\gamma$ which
  can, in principle, be removed by measuring the phase of
  $V_{cd}/V_{ud}$ independently.}. At a scale of order $m_b$ the two
transitions are then described by the leading nonleptonic weak
effective Hamiltonians~\cite{Buchalla:1995vs}
\begin{eqnarray}\label{Heffcu}
{\cal H}_{\bar c u}^{(0)} = 
\frac{G_F}{\sqrt2} V_{cb}^{\phantom{*}} V^*_{ud}\big[ C_1(\mu) Q_1^{\bar c u}
+ C_2(\mu) Q_2^{\bar c u}  \big], \\
\label{Heffuc}
{\cal H}_{\bar u c}^{(0)} = 
\frac{G_F}{\sqrt2} V_{ub}^{\phantom{*}} V^*_{cd}\big[ C_1(\mu) Q_1^{\bar u c}
+ C_2(\mu) Q_2^{\bar u c}  \big]
\end{eqnarray}
which involve the usual four-fermion operators defined by
\begin{align}
Q_1^{\bar c u}&= (\bar cb)_{V-A}(\bar du)_{V-A}, \qquad
Q_2^{\bar c u}= (\bar d b)_{V-A}(\bar c u)_{V-A}  , \label{Q12}\\
 Q_1^{\bar u c}&= (\bar ub)_{V-A}(\bar dc)_{V-A},\qquad
Q_2^{\bar u c}=(\bar d b)_{V-A}(\bar u c)_{V-A} . 
\end{align}
Here, $(\bar qq')_{V-A}$ denotes the left-handed structure $\bar
q\gamma^\mu(1-\gamma_5)q'$, for quark fields $q, q'$. The Wilson
coefficients, evaluated at a scale of the order of the $b$-quark mass
$\mu\sim m_b$, are given by $C_1(m_b)=1.10$ and $C_2(m_b)=-0.24$ at
leading-log order, for $m_b(m_b) =
4.163\,\text{GeV}$~\cite{Chetyrkin:2009fv} and the strong coupling
constant $\alpha_s(M_Z)=0.1184$~\cite{Beringer:1900zz}. The decay
amplitudes in Eq. \eqref{rB} are then given, at leading order in the
electroweak interactions, by
\begin{equation}\label{eqA}
A(B^- \to \bar D^0 \pi^-)=\langle \bar D^0 \pi^-| {\cal H}_{\bar u
  c}^{(0)}| B^-\rangle, \quad A(B^- \to D^0 \pi^-)=\langle D^0 \pi^-|
{\cal H}_{\bar c u}^{(0)}| B^-\rangle.
\end{equation}

\begin{figure}
    \centering
\includegraphics[width=0.3\columnwidth]{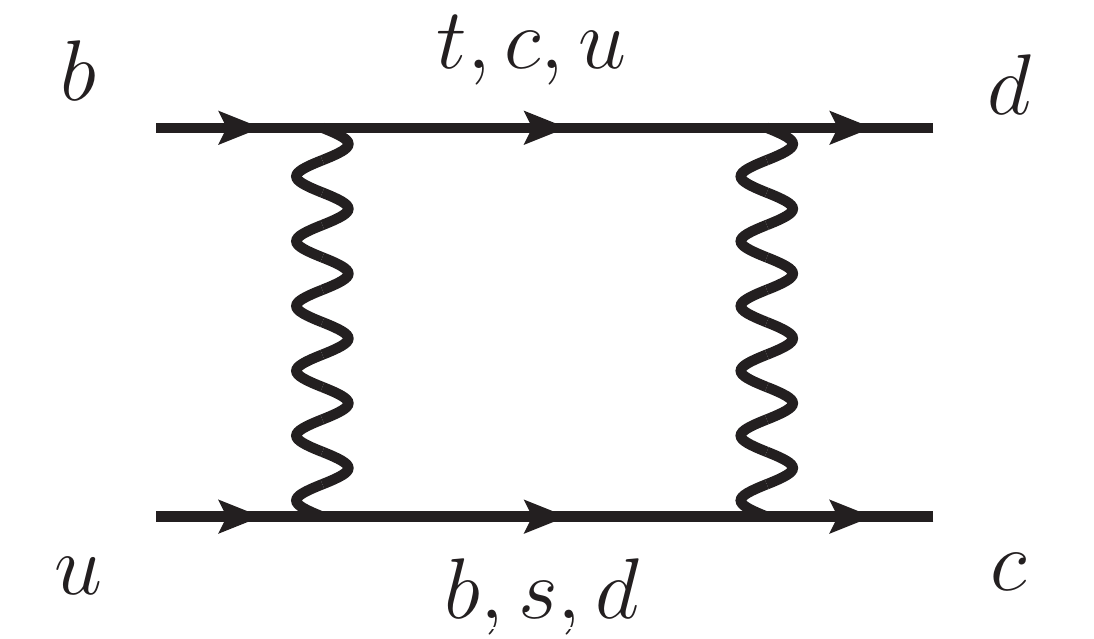}~~~~~~~~~~~~~~~~~~~
\includegraphics[width=0.3\columnwidth]{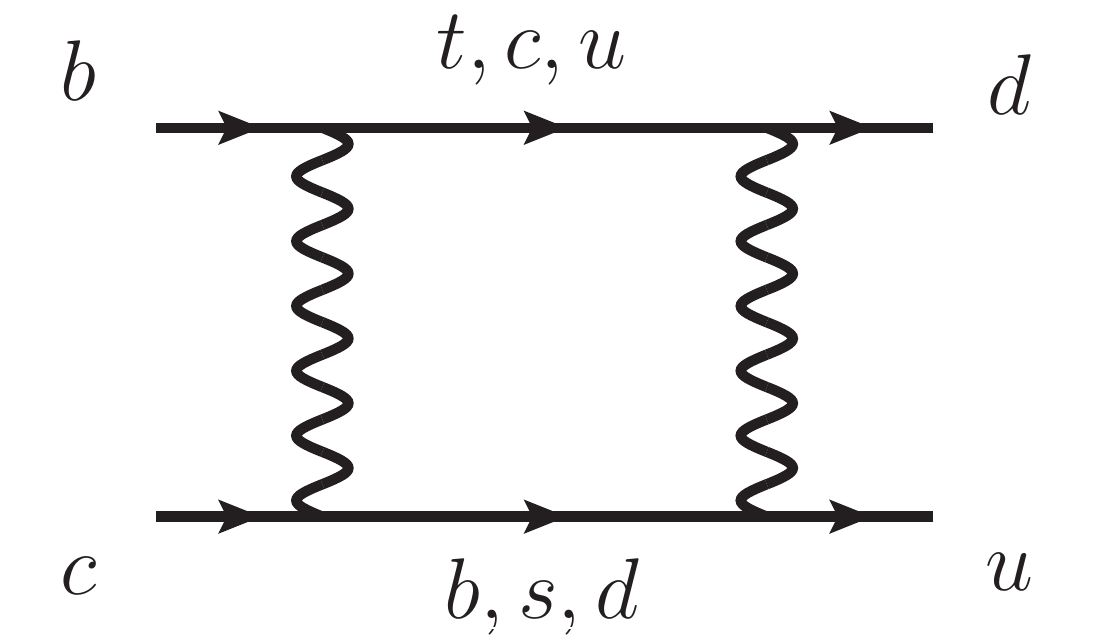}  
\caption{ The electroweak corrections to the $b\to c \bar u d$ and
  $b\to u \bar c d$ processes at order ${\mathcal O}(G_F^2)$. Curly
  lines represent $W$ bosons and the corresponding pseudo-Goldstone
  bosons.}
\label{fig:ewcorr}
\end{figure}

Electroweak corrections, of the order of ${\mathcal O}(G_F^2)$, to the
amplitudes will induce a shift $\delta \gamma^{D\pi}$ in the extracted
value of $\gamma$ if the ${\mathcal O}(G_F)$ and ${\mathcal O}(G_F^2)$
contributions differ in their weak phase. As argued
in~\cite{Brod:2013sga}, the only second-order weak corrections
to~\eqref{rB} and~\eqref{eqA} that need to be considered are those
arising from $W$ box diagrams that have a different CKM structure than
the corresponding tree amplitude, see Fig.~\ref{fig:ewcorr}. (Diagrams
with photon or $Z$-boson exchange do not lead to a different CKM
structure, whereas $W$ vertex corrections can be absorbed into a
universal renormalization of the CKM matrix elements.)

For instance, the CKM structures of the $b \to u \bar c d$ transition
(left diagrams in Fig.~\ref{fig:tree} and Fig.~\ref{fig:ewcorr}) are
given by $V_{ub}^{\phantom{*}} V^*_{cd}$ for the tree-level diagram
and $(V_{tb}^{\phantom{*}} V^*_{td})(V_{ub}^{\phantom{*}} V^*_{cb})$
for the box diagram. They differ in their weak phases and thus lead to
a shift in the extracted value of $\gamma$.

The $b\to c \bar u d$ transition receives a similar correction (right
diagrams in Fig.~\ref{fig:tree} and Fig.~\ref{fig:ewcorr}), with CKM
structures $V_{cb}^{\phantom{*}} V^*_{ud}$ at tree level and
$(V_{tb}^{\phantom{*}} V^*_{td})(V_{cb}^{\phantom{*}} V^*_{ub})$ for
the box diagram. The effects of these diagrams are CKM suppressed with
respect to the previous contribution by two orders of magnitude and
can be safely neglected.

To a very good approximation, the only effect of the box diagrams is
thus a correction to the Wilson coefficients in the effective
Hamiltonian~\eqref{Heffuc}. Keeping only the local parts of the box
diagrams we can write
\begin{equation}\label{H1}
{\cal H}_{\bar u c}^{(1)} = \frac{G_F}{\sqrt2} V_{ub}^{\phantom{*}} V^*_{cd}\big[
  \big(C_1(\mu)+\Delta C_1(\mu)\big) Q_1^{\bar u c} +
  \big(C_2(\mu)+\Delta C_2(\mu)\big) Q_2^{\bar u c} \big].
\end{equation}
The Wilson coefficients $C_{1,2}(\mu)$ are the same as in
Eqs.~\eqref{Heffcu} and~\eqref{Heffuc}, while $\Delta C_{1,2}(\mu)$
are corrections of ${\mathcal O}(G_F)$ relative to the tree-level
diagrams. They depend on the CKM elements and carry a weak phase
different from that of $C_{1,2}(\mu)$ (which are real in our
convention). While the precise absolute values of the Wilson
coefficients are irrelevant for the experimental analysis, where all
branching fractions and amplitude ratios are fitted from data, a
contribution with a relative weak phase will induce a shift in the
extracted value of $\gamma$.

To get a first estimate of the size of the effect we will perform a
matching calculation from the SM directly onto the weak effective
Hamiltonian where the $W$ boson, the top quark, and the bottom quark
have been integrated out simultaneously. To this end, we evaluate the
box diagrams in Fig.~\ref{fig:ewcorr} at $\mu\sim M_W$, treating the
top and bottom quarks as massive and all remaining quarks as massless,
and setting all external momenta to zero. Because of the
Glashow-Iliopoulos-Maiani mechanism acting on both the internal
up-quark and down-quark lines the result is proportional to $x_t y_b$,
where $x_t\equiv m_t^2/M_W^2, y_b\equiv m_b^2/M_W^2$, and we find for
the shift $\Delta C_2$ of the Wilson coefficient $C_2$ in
Eq.~\eqref{H1}
\begin{multline}\label{DeltaC2}
\Delta C_2 = -\sqrt{2} G_F \, \frac{
  M_W^2}{4\pi^2}\,\frac{V_{tb}^{\phantom{*}} V_{td}^*
  V_{cb}^*}{V_{cd}^*} \, \hat C(x_t,y_b) \\ = -\sqrt{2}G_F \, \frac{
  M_W^2}{4\pi^2} \left|\frac{V_{tb}V_{td}V_{cb}}{V_{cd}}\right|
e^{i\beta} \, \hat C(x_t,y_b)\,,
\end{multline}
with the CKM angle $\beta \equiv \arg(-V_{cd}^{\phantom{*}}
V_{cb}^*/V_{td}^{\phantom{*}} V_{tb}^*)$ and the loop function
\begin{equation}\label{eq:Chatfull}
\hat C(x_t,y_b) = \frac{x_t \, y_b}{8} \bigg[
  \frac{9}{(x_t-1)(y_b-1)} + \bigg(
  \frac{(x_t-4)^2}{(x_t-1)^2(x_t-y_b)} \log x_t + (x_t \leftrightarrow
  y_b) \bigg) \bigg] \, .
\end{equation}
The result of our calculation agrees with the corresponding loop
function extracted from~\cite{Inami:1980fz}. In this first estimate,
the shift of the Wilson coefficient $C_1$ is zero. Using the input
from~\cite{Beringer:1900zz} we find
\begin{equation}
\begin{split}\label{eq:unresum}
\Delta C_2 &= - (1.18 \pm 0.11) \cdot 10^{-7} \times e^{i\beta} \, ,
\end{split}
\end{equation}
where the shown error is dominated by the uncertainty on the CKM
elements $V_{tb}$, $V_{cb}$, and $V_{td}$.

\begin{figure}
    \centering
\includegraphics[width=0.28\columnwidth]{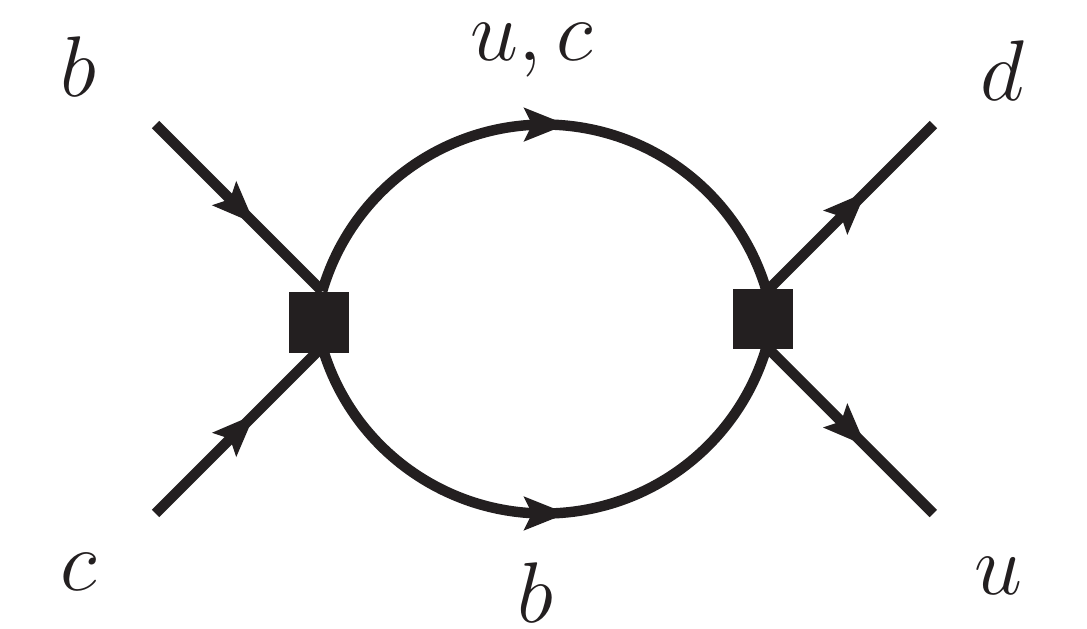}~~~
\includegraphics[width=0.28\columnwidth]{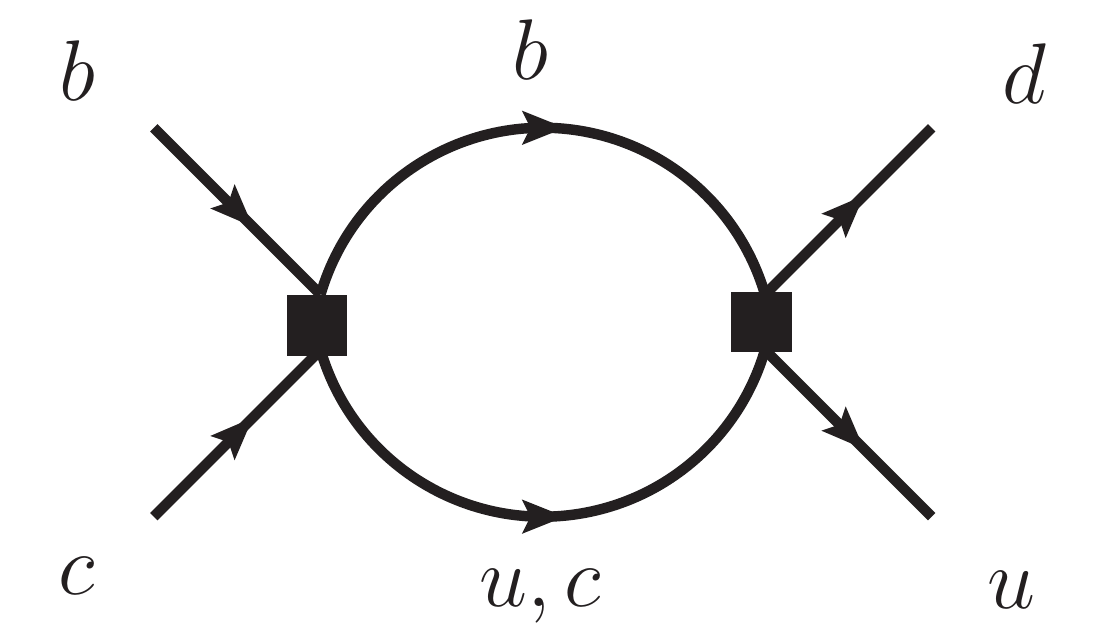}~~~ 
\includegraphics[width=0.28\columnwidth]{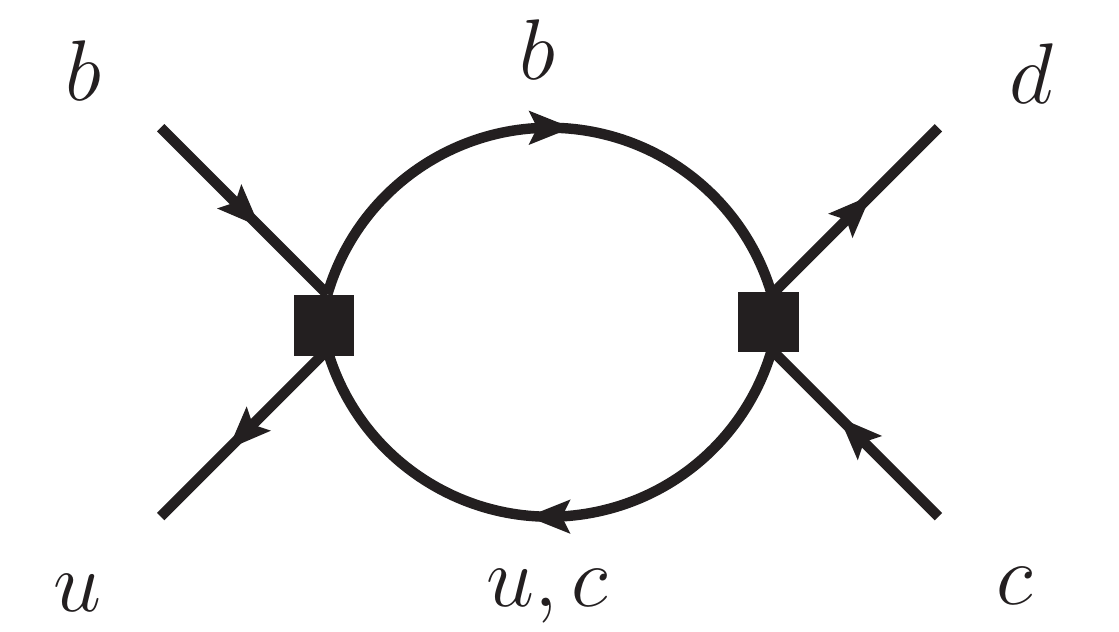}
    \caption{ The double insertion $T\{Q_1,Q_1\}$ (left) and
      $T\{Q_2,Q_2\}$ (middle and right), contributing to the mixing
      into $\tilde Q_2$.  }
\label{fig:matching2}
 \end{figure}

\begin{figure}
    \centering
\includegraphics[width=0.3\columnwidth]{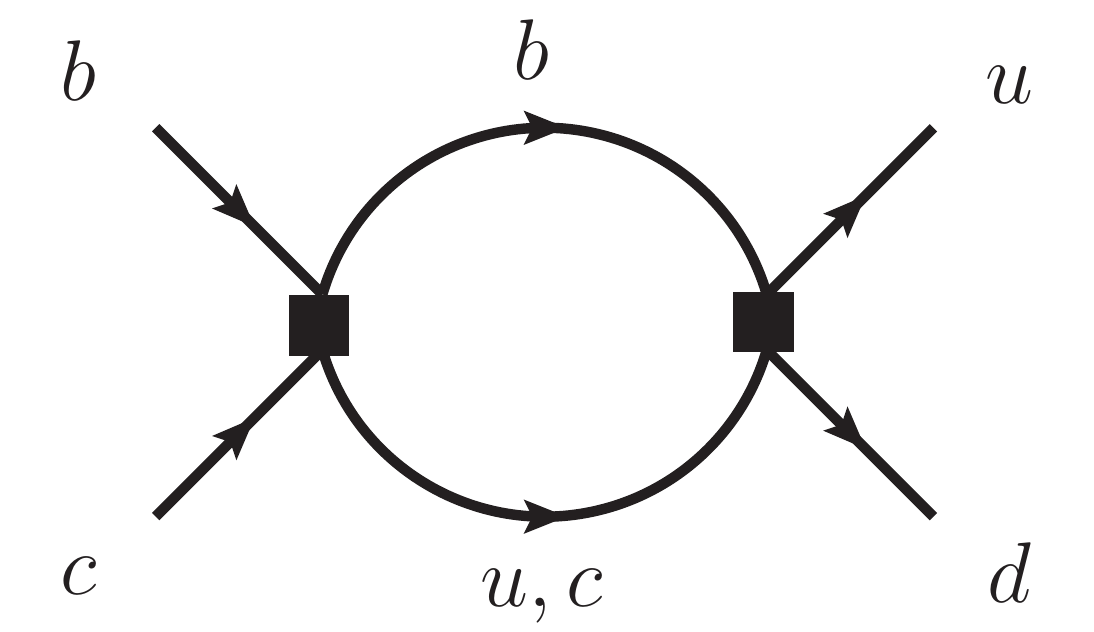}~~~~~~~~
\includegraphics[width=0.3\columnwidth]{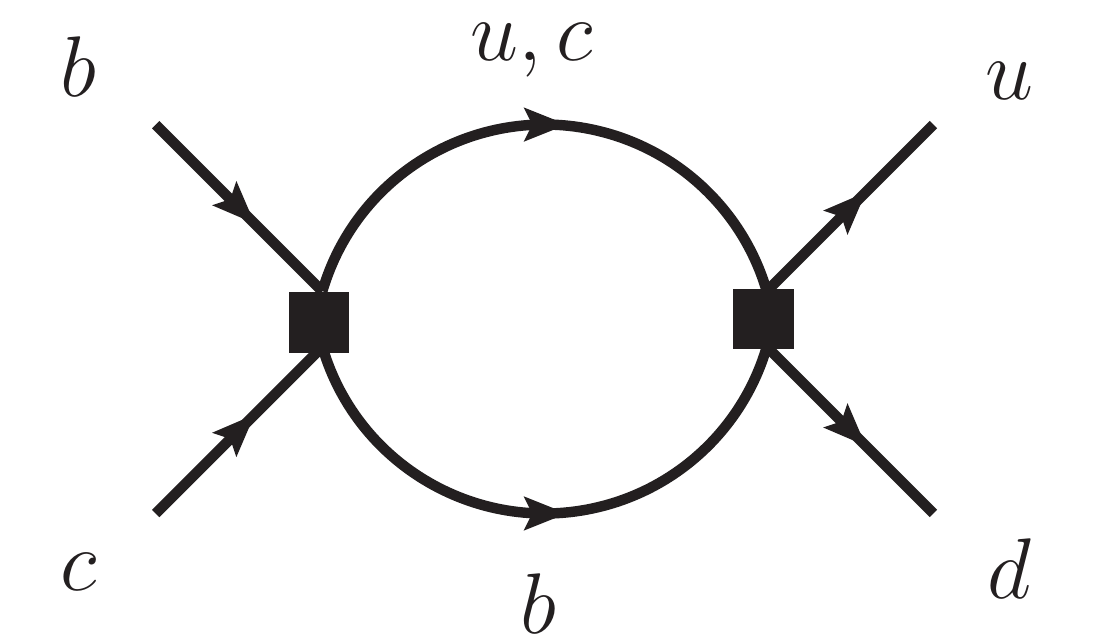} 
    \caption{ The double insertions $T\{Q_1,Q_2\}$ contributing to the
      mixing into the operator $\tilde Q_1$.}
\label{fig:matching3}
\end{figure}

The loop function $\hat C(x_t,y_b)$ is dominated by the term
proportional to $\log y_b$:
\begin{equation}\label{eq:Chat}
\hat C(x_t,y_b) = - 2 y_b \log y_b+{\mathcal O}(y_b)\,,
\end{equation}
where the subleading terms amount to a 10\% correction. In order to
capture also the leading QCD corrections, we now refine our analysis
and perform a resummation of the terms proportional to $\log y_b$ to
all orders in the strong coupling constant. To achieve this, we first
match the SM to the effective theory below the scale $\mu_W =
{\mathcal O}(M_W)$, where the top quark and the heavy gauge bosons are
integrated out, but the bottom quark is still a dynamical degree of
freedom. In fact, the matching correction at $\mu_W$ vanishes to
leading order. However, the renormalization-group (RG) running will
generate this term at the bottom-quark scale $\mu_b = {\mathcal
  O}(m_b)$ via bilocal insertions of the effective Hamiltonian
\begin{equation}\label{eq:H5}
\begin{split}
{\cal H}_{\rm eff}^{f=5} & = \frac{G_F}{\sqrt2} \sum_{\substack{u_{1,2}=u,c\\d_{1,2}=s,d,b}}
  V_{u_1d_2}^{\phantom{*}} V^*_{u_2d_1} \sum_{i,j = 1}^2 C_i(\mu) Z_{ij }Q_j^{(u_1
    d_2; d_1 u_2)} \\
& - 2G_F^2 V_{ub}^{\phantom{*}} V_{cd}^* \cdot \bigg| \frac{V_{tb} V_{td}
  V_{cb}}{V_{cd}} \bigg| e^{i\beta} \bigg[ \sum_{i,j,k=1}^2 C_i C_j
\hat Z_{ij,k} \tilde Q_k +\sum_{l,k=1}^2 \tilde C_l \tilde Z_{lk}
\tilde Q_k \bigg]  \, .
\end{split}
\end{equation} 
Here, $Z$ and $\hat Z$ are the renormalization constants for the local
and bilocal insertions, respectively. The first line in
Eq.~\eqref{eq:H5} contains the four-quark operators obtained by
integrating out the $W$ and $Z$ bosons. We denote them by
\begin{equation}
  Q_1^{(u_1 d_2;d_1 u_2)} = (\bar u_1 d_2)_{V-A} (\bar d_1
  u_2)_{V-A} \, , \quad Q_2^{(u_1 u_2;d_1 d_2)} = (\bar
  u_1 u_2)_{V-A} (\bar d_1 d_2)_{V-A}\, .   
\end{equation}
The second line in Eq.~\eqref{eq:H5} contains the operators
\begin{equation}
  \tilde Q_1 = \frac{m_b^2}{\mu^{2\epsilon} g_s^2} (\bar u b)_{V-A}
  (\bar d c)_{V-A} \, , \quad \tilde Q_2 =
  \frac{m_b^2}{\mu^{2\epsilon} g_s^2} (\bar d b)_{V-A}(\bar u c)_{V-A}
  \, . \label{tildeQ12}
\end{equation}
They arise as counterterms to the bilocal insertions and are thus
formally of dimension eight; this is made explicit by the $m_b^2$
prefactor. These operators have the same four-quark structure as the
leading-power operators $Q_{1,2}$. We neglect the six-quark operators
which arise from integrating out the $W$ boson and the top quark, as
they are suppressed by an additional factor of $1/M_W^2$.

To arrive at the CKM structure of the second line in Eq.~\eqref{eq:H5}
we note first that the two diagrams in Fig.~\ref{fig:matching2}
(right) have exactly the same phase as the corresponding tree-level
diagram, so we can drop them. For the remaining diagrams we use the
unitarity relation $V_{ub}^{\phantom{*}} V_{ud}^* +
V_{cb}^{\phantom{*}} V_{cd}^* = - V_{tb}^{\phantom{*}} V_{td}^*$,
combining pairs of diagrams with internal up and charm quarks as shown
in Fig.~\ref{fig:matching2} and~\ref{fig:matching3}, and then factor
out the tree-level coefficient $V_{ub}^{\phantom{*}} V_{cd}^*$.

The relevant diagrams in Fig.~\ref{fig:matching2}
and~\ref{fig:matching3} yield the following mixing (we use
$\hat\gamma_{i,j;k} = 2\hat Z_{i,j;k}$ and expand $\hat\gamma_{i,j;k}
= \frac{\alpha_s}{4\pi} \hat\gamma_{i,j;k}^{(0)} + \ldots$, where
$i,j$ denote the $Q_{1,2}$ insertions, and $k$ is the label of the
$\tilde Q_k$ operators):
\begin{equation}
\begin{split}
\hat\gamma_{1,1;2}^{(0)} = \hat\gamma_{2,2;2}^{(0)} = 
\hat\gamma_{1,2;1}^{(0)} = \hat\gamma_{2,1;1}^{(0)} = - 8 \, , 
\end{split}
\end{equation}
with all the remaining entries either vanishing or not contributing.
The value of the Wilson coefficients $\tilde C_k$ at the scale $m_b$
can now be calculated in complete analogy to the procedure in
Ref.~\cite{Brod:2013sga}, where we refer the interested reader for
details. Our final result, using $m_b(m_b) =
4.163\,\text{GeV}$~\cite{Chetyrkin:2009fv}, $\alpha_s(M_Z) =
0.1184$~\cite{Beringer:1900zz}, and employing
``RunDec''~\cite{Chetyrkin:2000yt} for the numerical running of the
strong coupling constant, is
\begin{equation}
\big(\tilde C_1(m_b), \tilde C_2(m_b)\big) =
(0.03, 0.31) \, .
\end{equation}

Finally, at the bottom-quark scale we need to match the matrix
elements of the two Hamiltonians~\eqref{eq:H5}
and~\eqref{Heffuc}. This will yield the leading $y_b$ behaviour with
resummed logarithms. We write the matrix elements as
\begin{equation}
\begin{split}\label{eq:matchlow}
\sum_{k}
\Delta C_k(\mu_b) \langle Q_k \rangle (\mu_b) = -
2\sqrt{2}G_F \left |
\frac{V_{tb}V_{td}V_{cb}}{V_{cd}} \right | e^{i\beta} \sum_{i=1,2}
\tilde C_i(\mu_b) \langle \tilde Q_i \rangle (\mu_b) \, ,
\end{split}
\end{equation}
where we expand $\Delta C_k = \frac{4\pi}{\alpha_s} \Delta C_k^{(0)} +
\ldots$ in such a way that the artificially inserted factor of
$1/g_s^2$ in the definition of $\tilde Q_k$~\eqref{tildeQ12} is
canceled. In Eq.~\eqref{eq:matchlow} we have dropped the double
insertions $\langle Q_i Q_j \rangle$ as they enter at higher order in
$\alpha_s$ and need not be calculated in our approximation. Therefore,
we effectively obtain the matching condition for the Wilson
coefficients of the local operators \eqref{H1} in the form
\begin{equation}
 \Delta C_k^{(0)}(\mu_b) = - 2 m_b^2  \frac{\sqrt{2}G_F}{16\pi^2} \left |
\frac{V_{tb}V_{td}V_{cb}}{V_{cd}} \right | e^{i\beta}  \tilde C_k^{(0)}(\mu_b) \, .
\end{equation}
Numerically, we find
\begin{equation}\label{DeltaC12-resum}
\Delta C_1 = - (1.14 \pm 0.10) \cdot 10^{-8} \times e^{i\beta} \, ,
\quad \Delta C_2 = - (1.09 \pm 0.09) \cdot 10^{-7} \times e^{i\beta}\, ;
\end{equation}
the quoted errors reflect the uncertainty in the electroweak input
parameters. This should be compared to the unresummed result
Eq.~\eqref{eq:unresum}: we see that, indeed, the RG running has
induced a nonzero correction to the Wilson coefficient $C_2$
in~\eqref{H1}. Moreover, also $C_1$ gets a small correction, in
contrast to the unresummed result. As a check of our calculation we
expand the solution of the RG equations about $\mu=M_W$ and recover
exactly the logarithm in Eq.~\eqref{eq:Chat},
\begin{equation}\label{DeltaC12-expanded}
\Delta C_1 = 0 \, , \quad \Delta C_2 \propto - \sqrt{2} G_F \,
\frac{M_W^2}{4\pi^2} ( - 2 y_b \log y_b) \, ,
\end{equation}
where we dropped the CKM factors.

\section{The induced shift in $\gamma$}
\label{sec:shift}

The imaginary part of the shift in the Wilson coefficients calculated
in the previous two sections induces a shift in $\gamma$ via a
modification of the ratio $r_B^{D\pi}$, Eq.~\eqref{rB}:
\begin{equation}\label{eq:rBDpi}
r_B^{D\pi} e^{i(\delta_B^{D\pi}-\gamma)}\to r_B^{D\pi} e^{i(\delta_B^{D\pi}-\gamma)}
\Big(1 + \frac{\Delta C_1}{C_1 + C_2 r_{A'}} + \frac{\Delta C_2}{C_1/r_{A'} + C_2} \Big)\,,
\end{equation}
where we expanded in the small corrections $\Delta C_1$, $\Delta C_2$
to linear order. The resulting shift in the extracted value of
$\gamma$ is
\begin{equation}\label{eq:gammacorr}
\delta \gamma^{D\pi} =  - \frac{{\rm Im}(\Delta C_1)}{C_1 + C_2 r_{A'}} -
\frac{{\rm Im}(\Delta C_2)}{C_1/r_{A'} + C_2}\, .
\end{equation}
To estimate its size we need to evaluate the amplitude ratio $r_{A'}$,
defined as
\begin{equation}\label{rA}
r_{A'}\equiv\frac{\langle \pi^- \bar D^0| Q_2^{\bar
    uc}|B^-\rangle}{\langle \pi^- \bar D^0|Q_1^{\bar uc}|B^-\rangle}\,.
\end{equation}
The amplitudes contain the $\bar D^0$ meson in the final state; this
is directly related to the fact that the electroweak corrections
affect only the numerator of the ratio~\eqref{rB}. By contrast, in the
case of $B \to D K$ only the denominator of the corresponding
amplitude ratio $r_B^{DK}$ is modified (the reason being the different
CKM structure of the $B \to D K$ modes).

Keeping in mind that the $D$ meson is much heavier than the pion we
see that both numerator and denominator in $r_{A'}$ are suppressed by
powers of $\Lambda_\text{QCD}/m_b$~\cite{Beneke:2000ry}. Using color
counting and neglecting annihilation topologies yields $r_{A'} \sim
N_c = 3$ as a naive estimate, with large uncertainties. A crude
numerical estimate treating both final-state particles as
light~\cite{Beneke:2001ev} and using an asymmetric $D$-meson wave
function~\cite{Beneke:2000ry} suggests that the annihilation
contribution is indeed negligible and that $r_{A'} \approx 1$.

Interestingly, for a value of $r_{A'} \approx 4.6$ the two terms in
the denominators in Eq.~\eqref{eq:gammacorr} cancel each other, so
that the electroweak correction to the ratio $r_B^{D\pi}$ could, in
principle, become arbitrarily large. The reason, of course, is that
this cancellation would imply the vanishing of $r_B^{D\pi}$. Ignoring
differences in the matrix elements related to the replacement of pions
by kaons, this would also imply the vanishing of the ratio $r_B^{DK}$,
in contradiction to the measured value (cf. the discussion below
Eq.~\eqref{rB}). A complete cancellation can thus be safely excluded,
although a more quantitative statement is difficult to obtain. To be
conservative we will take $r_{A'} = 4.5$ for our estimate of $\delta
\gamma^{D\pi}$. Using $\sin2\beta=0.682$~\cite{Beringer:1900zz} we
then obtain
\begin{equation}
\delta \gamma^{D\pi}\simeq 9.7 \cdot 10^{-6}\, \text{(unresummed)}
\,, \quad \delta \gamma^{D\pi}\simeq 9.2 \cdot 10^{-6}\,
\text{(resummed)} \,.
\end{equation}
Large uncertainties are associated with these numbers due to the
poorly known value of $r_{A'}$ and missing nonlocal contributions, but
it seems very unlikely that the shift in $\gamma$ exceeds $\big|\delta
\gamma^{D\pi}/\gamma\big|\lesssim 10^{-4}$. Note that for values of
$r_{A'} \lessapprox 3$ the shift $\big|\delta
\gamma^{D\pi}/\gamma\big|$ drops below $10^{-6}$ . On the other hand,
considerable fine tuning would be required for an almost complete
cancellation of the denominators in Eq.~\eqref{eq:gammacorr}. For
instance, to find $|\delta \gamma^{D\pi}/\gamma|$ larger than
$10^{-3}$ would require a tuning of $r_{A'}$ of the order of
$10^{-4}$.

\section{Summary and Conclusion}
\label{Conclusions}

The determination of the CKM phase $\gamma$ from tree-level decays is
theoretically exceptionally clean, as all necessary branching
fractions and amplitude ratios can be obtained from experimental
data. In the SM, the only shift in $\gamma$ is induced by electroweak
corrections to the effective Hamiltonian that carry a weak phase
relative to the leading contributions. In this letter we have
estimated the shift for the extraction of $\gamma$ from the $B \to D
\pi$ decay modes. We calculated the electroweak corrections in two
ways, first integrating out the bottom quark together with the top
quark and the $W$ boson, then also summing leading QCD logs of
$m_b/M_W$ in a two-step matching procedure.

Interestingly, the different CKM structure compared to the $B \to D K$
modes could lead to a moderately large shift in $\gamma$ via an
approximate cancellation of hadronic matrix elements. Whereas these
matrix elements are hard to estimate, we find that without large
accidental fine tuning the expected shift in $\gamma$ is very unlikely
to exceed
\begin{equation}
\big|\delta \gamma^{D\pi}/\gamma\big|\lesssim 10^{-4}\,.
\end{equation}
A better estimate of the hadronic matrix elements seems worthwile and
could reduce this uncertainty.

{\bf Acknowledgements:} The author would like to thank Y. Grossman,
A. Lenz, M. Neubert, M. Savastio, and S. Turczyk for discussions and
J. Zupan for discussions and valuable comments on the manuscript. The
research of JB is supported by the ERC Advanced Grant EFT4LHC of the
European Research Council and the Cluster of Excellence Precision
Physics, Fundamental Interactions and Structure of Matter (PRISMA-EXC
1098).


\begin{thebibliography}{99}
%
%

\bibitem{Bigi:1981qs} 
  I.~I.~Y.~Bigi and A.~I.~Sanda,
  Nucl.\ Phys.\ B {\bf 193}, 85 (1981).

\bibitem{Gronau:1990ra}
  M.~Gronau and D.~London.,
  Phys.\ Lett.\ B {\bf 253}, 483 (1991).

\bibitem{Gronau:1991dp}
M.~Gronau and D.~Wyler,
Phys.\ Lett.\ B {\bf 265}, 172 (1991).

\bibitem{Atwood:1996ci}
D.~Atwood, I.~Dunietz and A.~Soni,
Phys.\ Rev.\ Lett.\  {\bf 78}, 3257 (1997);
D.~Atwood, I.~Dunietz and A.~Soni,
Phys.\ Rev.\ D {\bf 63}, 036005 (2001).

\bibitem{Giri:2003ty}
A.~Giri, Y.~Grossman, A.~Soffer and J.~Zupan,
Phys.\ Rev.\ D {\bf 68}, 054018 (2003).

\bibitem{Grossman:2002aq}
Y.~Grossman, Z.~Ligeti and A.~Soffer,
Phys.\ Rev.\ D {\bf 67}, 071301 (2003).

\bibitem{Bondar:2005ki}
  A.~Bondar and A.~Poluektov,
  Eur.\ Phys.\ J.\ C {\bf 47}, 347 (2006)
  [arXiv:hep-ph/0510246].

\bibitem{Kayser:1999bu}
B.~Kayser and D.~London,
Phys.\ Rev.\ D {\bf 61}, 116013 (2000).
D.~Atwood and A.~Soni,
Phys.\ Rev.\ D {\bf 68}, 033009 (2003);
  R.~Fleischer,
  Phys.\ Lett.\ B {\bf 562}, 234 (2003)
  [arXiv:hep-ph/0301255];
  D.~Atwood and A.~Soni,
  Phys.\ Rev.\ D {\bf 68}, 033009 (2003)
  [arXiv:hep-ph/0206045];
R.~Aleksan, I.~Dunietz and B.~Kayser,
Z.\ Phys.\  C {\bf 54} (1992) 653.

\bibitem{Gronau:2004gt}
  M.~Gronau, Y.~Grossman, N.~Shuhmaher, A.~Soffer and J.~Zupan,
  Phys.\ Rev.\ D {\bf 69}, 113003 (2004)
  [arXiv:hep-ph/0402055].

\bibitem{Aleksan:2002mh}
R.~Aleksan, T.~C.~Petersen and A.~Soffer,
Phys.\ Rev.\ D {\bf 67}, 096002 (2003).
M.~Gronau,
Phys.\ Lett.\ B {\bf 557}, 198 (2003);
D.~Atwood and A.~Soni,
Phys.\ Rev.\ D {\bf 68}, 033003 (2003).

\bibitem{Gershon:2009qr}
T.~Gershon and A.~Poluektov,
Phys.\ Rev.\  D {\bf 81} (2010) 014025.

\bibitem{Gershon:2008pe}
T.~Gershon,
Phys.\ Rev.\  D {\bf 79} (2009) 051301
[arXiv:0810.2706 [hep-ph]].

\bibitem{Gershon:2009qc}
T.~Gershon and M.~Williams,
Phys.\ Rev.\  D {\bf 80} (2009) 092002
[arXiv:0909.1495 [hep-ph]].

\bibitem{Bondar:2004bi}
A.~Bondar and T.~Gershon,
Phys.\ Rev.\  D {\bf 70} (2004) 091503
[arXiv:hep-ph/0409281].

\bibitem{Sinha:2004ct}
N.~Sinha,
Phys.\ Rev.\  D {\bf 70} (2004) 097501
[arXiv:hep-ph/0405061].

\bibitem{Zupan:2007zz}
J.~Zupan,
Nucl.\ Phys.\ Proc.\ Suppl.\  {\bf 170} (2007) 65;
arXiv:hep-ph/0410371;
chapter 8 of 
M.~Antonelli {\it et al.},
Phys.\ Rept.\  {\bf 494} (2010) 197
[arXiv:0907.5386 [hep-ph]].

\bibitem{Aaij:2013zfa} 
  R.~Aaij {\it et al.}  [LHCb Collaboration],
  Phys.\ Lett.\ B {\bf 726}, 151 (2013)
  [arXiv:1305.2050 [hep-ex]].

\bibitem{LHCbckm2014}
  The LHCb Collaboration, 
  LHCb-CONF-2014-004

\bibitem{Libby:2010nu}
J.~Libby {\it et al.} [CLEO Collaboration],
Phys.\ Rev.\ D {\bf 82} (2010) 112006
[arXiv:1010.2817 [hep-ex]].

\bibitem{Aihara:2012aw}
H.~Aihara {\it et al.} [Belle Collaboration],
Phys.\ Rev.\ D {\bf 85} (2012) 112014
[arXiv:1204.6561 [hep-ex]].

\bibitem{Aaij:2014uva} 
  R.~Aaij {\it et al.}  [LHCb Collaboration],
  JHEP {\bf 1410}, 97 (2014)
  [arXiv:1408.2748 [hep-ex]].

\bibitem{Silva:1999bd}
  J.~P.~Silva and A.~Soffer,
  Phys.\ Rev.\ D {\bf 61}, 112001 (2000)
  [arXiv:hep-ph/9912242].

\bibitem{Bondar:2010qs}
A.~Bondar, A.~Poluektov and V.~Vorobiev,
Phys.\ Rev.\  D {\bf 82} (2010) 034033
[arXiv:1004.2350 [hep-ph]].

\bibitem{Grossman:2005rp}
  Y.~Grossman, A.~Soffer and J.~Zupan,
  Phys.\ Rev.\ D {\bf 72}, 031501 (2005)
  [arXiv:hep-ph/0505270].

\bibitem{Rama:2013voa}
M.~Rama,
arXiv:1307.4384 [hep-ex].

\bibitem{Gronau:2007bh}
M.~Gronau, Y.~Grossman, Z.~Surujon and J.~Zupan,
Phys.\ Lett.\  B {\bf 649} (2007) 61
[arXiv:hep-ph/0702011].

\bibitem{Wang:2012ie} 
  W.~Wang,
  Phys.\ Rev.\ Lett.\  {\bf 110}, 061802 (2013)
  [arXiv:1211.4539 [hep-ph]].

\bibitem{Martone:2012nj} 
  M.~Martone and J.~Zupan,
  Phys.\ Rev.\ D {\bf 87}, 034005 (2013)
  [arXiv:1212.0165 [hep-ph]].

\bibitem{Bhattacharya:2013vc} 
  B.~Bhattacharya, D.~London, M.~Gronau and J.~L.~Rosner,
  Phys.\ Rev.\ D {\bf 87}, 074002 (2013)
  [arXiv:1301.5631 [hep-ph]].

\bibitem{Bondar:2013jxa} 
  A.~Bondar, A.~Dolgov, A.~Poluektov and V.~Vorobiev,
  arXiv:1303.6305 [hep-ph].

\bibitem{Grossman:2013woa} 
  Y.~Grossman and M.~Savastio,
  JHEP {\bf 1403}, 008 (2014)
  [arXiv:1311.3575 [hep-ph]].

\bibitem{Brod:2014bfa} 
  J.~Brod, A.~Lenz, G.~Tetlalmatzi-Xolocotzi and M.~Wiebusch,
  arXiv:1412.1446 [hep-ph].

\bibitem{Brod:2013sga} 
  J.~Brod and J.~Zupan,
  JHEP {\bf 1401}, 051 (2014)
  [arXiv:1308.5663 [hep-ph]].

\bibitem{Buchalla:1995vs}
G.~Buchalla, A.~J.~Buras and M.~E.~Lautenbacher,
Rev.\ Mod.\ Phys.\ {\bf 68} (1996) 1125
[arXiv:hep-ph/9512380].

\bibitem{Chetyrkin:2009fv} 
  K.~G.~Chetyrkin, J.~H.~Kuhn, A.~Maier, P.~Maierhofer, P.~Marquard, M.~Steinhauser and C.~Sturm,
  Phys.\ Rev.\ D {\bf 80}, 074010 (2009)
  [arXiv:0907.2110 [hep-ph]].

\bibitem{Beringer:1900zz} 
  J.~Beringer {\it et al.}  [Particle Data Group Collaboration],
  Phys.\ Rev.\ D {\bf 86}, 010001 (2012).

\bibitem{Inami:1980fz}
T.~Inami and C.~S.~Lim,
Prog.\ Theor.\ Phys.\ {\bf 65} (1981) 297
[Erratum-ibid.\ {\bf 65} (1981) 1772].

\bibitem{Chetyrkin:2000yt} 
  K.~G.~Chetyrkin, J.~H.~Kuhn and M.~Steinhauser,
  Comput.\ Phys.\ Commun.\  {\bf 133}, 43 (2000)
  [hep-ph/0004189].

\bibitem{Beneke:2000ry} 
  M.~Beneke, G.~Buchalla, M.~Neubert and C.~T.~Sachrajda,
  Nucl.\ Phys.\ B {\bf 591}, 313 (2000)
  [hep-ph/0006124].

\bibitem{Beneke:2001ev} 
  M.~Beneke, G.~Buchalla, M.~Neubert and C.~T.~Sachrajda,
  Nucl.\ Phys.\ B {\bf 606}, 245 (2001)
  [hep-ph/0104110].

\end{thebibliography}
\end{document}